\documentclass{siamltex}
\usepackage{color}
\usepackage{amsfonts,amsmath}
\usepackage{graphicx}
\usepackage{tikz}
\usepackage{algorithm}
\usepackage{algorithmic}
\usepackage{subfigure}
\usepackage{booktabs}
\usepackage{wasysym}
\usepackage{multirow}
\usepackage{enumerate}

\usetikzlibrary{arrows,positioning}
\tikzstyle{line} = [draw, -]

\newcommand{\ninv}{n_\mathit{inv}}

\newcounter{confcounter}
\newcommand{\conflabel}[1]{\refstepcounter{confcounter} \label{#1}}
 
\title{An Adaptive Aggregation Based Domain Decomposition Multilevel Method for
  the Lattice Wilson Dirac Operator: Multilevel Results
  \thanks{This work was partially funded by the Deutsche
    Forschungsgemeinschaft (DFG) Transregional Collaborative Research
    Centre 55 (SFB/TRR55)}} 

\author{A. Frommer\thanks{Department of Mathematics, Bergische
    Universit\"at Wuppertal, 42097 Germany, {\tt
      \{frommer,kkahl,leder,rottmann\} @math.uni-wuppertal.de}.}
  \and K. Kahl\footnotemark[2]
  \and S. Krieg\thanks{Department of Physics, Bergische
    Universit\"at Wuppertal, 42097 Germany and J\"ulich
    Supercomputing Centre, Forschungszentrum J\"ulich, 52428
    J\"ulich, Germany, {\tt s.krieg@fz-juelich.de}.}
  \and B.~Leder\footnotemark[2]
  \and M.~Rottmann\footnotemark[2]
}

\begin{document}

\maketitle

\begin{abstract}
In lattice QCD computations a substantial amount of work is spent in
solving linear systems arising in Wilson's discretization of the Dirac
equations. We show first numerical results of the extension of the
two-level DD-$\alpha$AMG method to a true multilevel method based on our parallel
\texttt{MPI}-\texttt{C} implementation. Using additional levels
pays off, allowing to cut down the core minutes spent on one system
solve by a factor of approximately $700$ compared to standard Krylov subspace
methods and yielding another speed-up of a factor of $1.7$ over the
two-level approach.
\end{abstract}



\section{Introduction}\label{sec:introduction}
In~\cite{Frommer:2013fsa} we recently proposed an adaptive aggregation
based domain decomposition
two-level (``DD-$\alpha$AMG'') method to solve linear systems 
\begin{equation}\label{eq:wilsondirac}
  D z = b
\end{equation} arising in Wilson's discretization of the Dirac
equations. Solving these systems makes up a large part of the compute time
spent in lattice QCD simulations~\cite{PRACE:ScAnnRep12,PRACE:ScC12}. These 
are among the most demanding in computational science, and thus triggered
intense research activity in the construction of
suitable preconditioners and, consequently, more efficient solvers for these systems in recent
years~\cite{MGClark2010_1,MGClark2007,Frommer:2013fsa,Luescher2007,MGClark2010_2}.

Our two-level method combines a multiplicative Schwarz
method (SAP) as the smoother with an aggregation based coarse-level
correction, components which were also used in the construction of an ``inexact
deflation'' approach in~\cite{Luescher2007}. In contrast to the
approach developed in~\cite{Luescher2007}, the two-level method
from~\cite{Frommer:2013fsa} arranged these ingredients
in a ``multigrid'' fashion, and thus the coarse-level system needed to be 
solved only to very low accuracy in each iteration. This yielded the 
fastest run times for the two-level method and now opens the path for a
 true multilevel method which we present in this paper.

Another multilevel approach for~\eqref{eq:wilsondirac}, which also uses an aggregation based 
coarse-level correction but a different, non-stationary smoothing iteration, has been 
developed
in~\cite{MGClark2010_1,MGClark2007,MGClark2010_2}. Significant
speed-ups over traditional Krylov subspace methods were reported for
this approach. We replace the smoother used in ~\cite{MGClark2010_1,MGClark2007,MGClark2010_2} 
by SAP in order to be able to benefit from data locality and improve the strong scaling.
In addition, the new setup
routine for our adaptive multilevel domain decomposition  approach differs from the one
used in~\cite{MGClark2010_2} in an important aspect: By combining the
``inverse-iteration''-type approach from~\cite{Luescher2007} with
a bootstrap-type approach from~\cite{KahlBootstrap} we are able to generate
appropriate test vectors more efficiently.

Experiments reported in~\cite{Frommer:2013fsa}
show significant speed-up of the two-level DD-$\alpha$AMG method over conventional
Krylov-subspace methods and notable speed-up over the
other hierarchical preconditioners mentioned above. Without going into detail
about the specific implementation of the method
in~\cite{Frommer:2013fsa}, we note that its error propagator is---as for many other two-level
approaches---of the generic form 
\begin{equation*}
  E_{2g} = (I-MD)^{\nu}(I-P D_{c}^{-1} R D)(I-MD)^{\mu}.
\end{equation*}
Here, $M$ denotes the smoother, $\mu$
and $\nu$ are the number of pre- and post-smoothing iterations, and $P$ and $R$
the interpolation and restriction operators, respectively. Note, that in
practice $D_{c}^{-1}$ can and will be approximated to low accuracy. 

The Wilson Dirac matrix, $D$, is $\gamma_5$-symmetric, i.e.,
$D\Gamma_5 = \Gamma_5 D^\dagger$ with $\Gamma_5$
acting as the identity on spins 1 and 2 and the negative identity on spins 3 and 4, see~\cite{Frommer:2013fsa}. 
We chose a $\gamma_5$-symmetry preserving Galerkin construction of
the coarse-level operator so that we have $R :=
P^{\dagger}$, where $P^{\dagger}$ denotes the conjugate transpose of
$P$, and $D_{c} := P^{\dagger}DP.$ By constructing the interpolation
$P$ to group only spin $1$ and $2$ as well as spin $3$ and $4$ variables, the
$\gamma_5$-symmetry of $D$ is preserved on the coarse-level, i.e., $D_{c}$ is again
$\gamma_5$-symmetric. The transition to a multilevel method is in principle straight-forward: one simply 
uses another two-level ansatz for the solution of the linear
system involving $D_{c}$ and applies this construction principle
recursively until a level is reached where a direct, ``exact'' solution of the 
coarse-level system is feasible.

The remainder of this note is organized
as follows. In Section $2$ we introduce some notation for the multilevel method
and give details about the cycling strategy used and the setup
employed. Thereafter we present extensive numerical studies conducted
with a three- and four-level method in Section $3$, showing that the
speed-ups anticipated in \cite{Frommer:2013fsa} are achieved in practice. 
Finally, we give some concluding remarks and an outlook on possible future progress in Section $4$.

\section{Domain Decomposition Adaptive Algebraic Multigrid}\label{sec:multilevel}
Our multilevel extension of the two level-approach from~\cite{Frommer:2013fsa}
combines the two same components, namely a multiplicative Schwarz method
(SAP)~\cite{Luescher2003,BFSmith_PEBjorstad_WDGropp_1996a} as the
smoother and a $\gamma_5$-symmetry preserving aggregation based
interpolation~\cite{MGClark2010_1,MGClark2007,Brezina2005,Frommer:2013fsa,MGClark2010_2}, on every level. 
This means that the smoother as well as the interpolation together with the associated 
coarse-level correction are of the same type on all levels of the hierarchy. 
As in the two-level case, and for the same reasons (cf.~\cite{Frommer:2013fsa}), the multilevel method is used as a preconditioner to a flexible
Krylov subspace method, e.g., FGMRES (see \cite{Saad:2003:IMS:829576}).

To be more specific, let $L$ denote the number of levels to be used in the
hierarchy and denote $D_{1} := D$. Then the setup of our method
constructs interpolation operators
$P_{\ell}$ for $\ell = 1,\ldots,L-1$, which transfer information from
level $\ell+1$ to level $\ell$, and computes
coarse-level operators $D_{\ell+1} =
(P_{\ell})^{\dagger}D_{\ell}P_{\ell}$. Given an SAP smoother, represented by its 
error propagation operator $I-M_{\ell}D_{\ell}$, on
every level as well, the simplest multilevel cycling strategy that can
be employed is the V-cycle illustrated in Algorithm~\ref{alg:Vcycle}. Here, 
on each level only one recursive call is used in-between pre- and
post-smoothing iterations.

\begin{algorithm}[htb]
  \caption{$z_{\ell} = \text{V-Cycle}(\ell,b_{\ell})$}\label{alg:Vcycle}
  \begin{algorithmic}[1]
    \IF{ $\ell = L$ }
      \STATE $z_\ell \leftarrow D_\ell^{-1} b_\ell$
    \ELSE
      \STATE $z_{\ell} = 0$
      \FOR{$i=1$ to $\mu$}
	\STATE $z_{\ell} \leftarrow z_{\ell} + M_{\ell} (b_\ell-D_\ell z_\ell)$
      \ENDFOR
      \STATE $b_{\ell+1} \leftarrow P_\ell^\dagger ( b_\ell - D_\ell z_\ell )$
      \STATE $z_{\ell+1} \leftarrow \text{V-Cycle}( \ell+1, b_{\ell+1} ) $\label{alg:line:recurrence}
      \STATE $z_{\ell}   \leftarrow z_{\ell} + P_\ell z_{\ell+1} $
      \FOR{$i=1$ to $\nu$}
	\STATE $z_{\ell}   \leftarrow z_{\ell} + M_{\ell} ( b_\ell - D_\ell z_\ell )$
      \ENDFOR
    \ENDIF  
  \end{algorithmic}
\end{algorithm}
 
\subsection{Multilevel K-Cycles}
Numerical tests with very large configurations and small quark masses
have shown that a simple V-cycle is often not the ideal choice in
terms of solver performance.
Thus we consider using a more elaborate cycling strategy, the K-cycle suggested in \cite{Notay2007}, in our
method. Instead of only one recursive call of the coarse-level solver,
a K-cycle optimally recombines several coarse-level solves. 
More precisely, on every level $\ell$ we
approximate the solution of the coarse-level 
system by a few iterations of a
flexible Krylov subspace method, which in turn is preconditioned
by the K-cycle multilevel method from level $\ell+1$ to $L$.
In here, we deviate from the approach in~\cite{Notay2007} by using a
stopping criterion based on the reduction of the associated residual
rather than a fixed number of iterations. We give the specific choice
of the stopping criterion used in our implementation in Section~\ref{sec:results}.

The K-cycle is illustrated in Algorithm~\ref{alg:Kcycle}. For a fixed number
of iterations it can be regarded  as a standard W-cycle; see e.g.~\cite{UTrottenberg_etal_2001},
with adaptive re-weighting of the approximate solutions after each recursion.

\begin{algorithm}[htb]
  \caption{$z_{\ell} = \text{K-Cycle}(\ell,b_{\ell})$}\label{alg:Kcycle}
  \begin{algorithmic}
    \STATE perform Algorithm~\ref{alg:Vcycle} with line~\ref{alg:line:recurrence} replaced by 
    \STATE $z_{\ell+1} \leftarrow \text{FGMRES for matrix $D_{\ell+1}$ and r.h.s.\ $b_{\ell+1}$, preconditioned with K-Cycle}( \ell+1, b_{\ell+1} ) $
  \end{algorithmic}
\end{algorithm}

\subsection{Multilevel Setup}\label{sec:setup}
For the construction of the aggregation based $\gamma_5$-symmetry preserving
interpolation operators $P_\ell$, and with them the coarse-level
operators $D_{\ell+1}$, we have extended our setup from~\cite{Frommer:2013fsa} to a multilevel
setup. In order to preserve the $\gamma_5$-symmetry on all
levels, we use a block-spin structure for the interpolation operators
on all levels. The setup process that we found to work best in practice is divided into
two phases: \bigskip

\begin{enumerate}[$1.$]
\item An initial phase given as Algorithm~\ref{alg:setup_init} which
constructs an initial multilevel hierarchy solely based on the
smoothing iteration starting with random test vectors.\medskip
\item An iterative phase given in Algorithm~\ref{alg:setup_iter} and
  illustrated in Fig.~\ref{fig:3lvlsetup}, where the current multilevel method
  is used to update and improve the multilevel hierarchy by generating improved test vectors.
\end{enumerate}\bigskip

Again we give our specific choices for the various parameters in the
setup in Section~\ref{sec:results}.
\begin{algorithm}
  \caption{initial\_setup\_phase($\ell$)}\label{alg:setup_init}
  \begin{algorithmic}[1]
    \IF{ $\ell = 0$ }
      \STATE Let $v^{(1)}_\ell,\ldots,v^{(N)}_\ell$  be $N$ random test vectors
    \ELSE
      \FOR{$j=1$ to $N$}
	\STATE $v^{(j)}_\ell \leftarrow P^\dagger_{\ell-1} v^{(j)}_{\ell-1} $ \COMMENT{restricted test vectors from previous level}
      \ENDFOR
    \ENDIF
    \FOR{$\eta=1$ to $3$}
      \FOR{$j=1$ to $N$}
	\STATE $x = 0$
	\FOR{$i=1$ to $\eta$}
	  \STATE $x \leftarrow x + M_\ell ( v^{(j)}_\ell - D_\ell x )$ \COMMENT{$M_\ell$ smoother for system with matrix $D_\ell$}
	\ENDFOR
	\STATE $v^{(j)} = x$
      \ENDFOR
    \ENDFOR
    \STATE construct $P_\ell$ and set $D_{\ell+1}= P_\ell^\dagger D_\ell P_\ell$
    \IF{ $\ell < L-1$ }
      \STATE initial\_setup\_phase($\ell+1$) \COMMENT{perform Algorithm~\ref{alg:setup_init} on next level}
    \ENDIF
  \end{algorithmic}
\end{algorithm}
\begin{algorithm}
  \caption{iterative\_setup\_phase($\ell$,$i$)}\label{alg:setup_iter}
  \begin{algorithmic}[1]
    \IF{ $\ell < L$ }
      \FOR{$j=1$ to $N$}
	\STATE $z_\ell \leftarrow \text{K-Cycle}( \ell, v^{(j)}_{\ell-1} )$
	\FOR{ $l=\ell$ to $L-1$ }
	  \STATE $v^{(j)}_l = z_l/||z_l||$ \COMMENT{update test vectors with the iterates of each level}
	\ENDFOR
      \ENDFOR
      \FOR{ $l = \ell,\ldots,L-1$ }
	\STATE construct $P_l$ and $D_{l+1}$
      \ENDFOR
      \FOR{$q=1$ to i}
	\STATE iterative\_setup\_phase($\ell+1,q$) \COMMENT{perform Algorithm~\ref{alg:setup_iter} on next level}
      \ENDFOR
    \ENDIF
  \end{algorithmic}
\end{algorithm}

\begin{figure}[htb]
  \begin{center}
    \resizebox{\textwidth}{!}{\begin{tikzpicture}[->,>=stealth',shorten >=1pt,auto,node distance=2.15cm,
  thick,main node/.style={circle,fill=black!20,draw,font=\sffamily\Large\bfseries}]

  \node[main node] (1)                      {$1$};
  \node[align=center] (20) [above of=1] {$\ell=1$, begin of the\\$i$-th iteration,\\current test vector\\as right hand side};
  \node[main node] (2)  [below right of=1]  {$2$};
  \node[main node] (3)  [below right of=2]  {$3$};
  \node[align=center] (30) [below of=3] {solve};
  \node[main node] (4)  [above right of=3]  {$2$};
  \node[align=center] (40) [above of=4] {smooth};

  \node[main node] (5)  [      right of=4]  {$2$};
  \node[main node] (6)  [below right of=5]  {$3$};
  \node[align=center] (31) [below of=6] {solve};
  \node[main node] (7)  [above right of=6]  {$2$};
  \node[align=center] (41) [above of=7] {smooth};
  \node[main node] (8)  [above right of=7]  {$1$};
  \node[align=center] (21) [above of=8] {smooth,\\re-build interpolations\\and coarse operators};
  
  \node[main node] (9)  [below right of=8]  {$2$};
  \node[align=center] (22) [above of=9] {proceed on\\next level};
  \node[main node] (10) [below right of=9]  {$3$};
  \node[align=center] (32) [below of=10] {solve};
  \node[main node] (11) [above right of=10] {$2$};
  \node[align=center] (23) [above of=11] {smooth,\\re-build interpolations\\and coarse operators};
  
  \node[main node] (12) [      right of=11] {$2$};
  \node[main node] (13) [below right of=12] {$3$};
  \node[align=center] (33) [below of=13] {solve};
  \node[main node] (14) [above right of=13] {$2$};

  \node[align=center] (24) [above of=14] {smooth,\\re-build interpolations\\and coarse operators};

  \path[every node/.style={font=\sffamily\small}]
    (1)  edge node [left]  {restrict} (2)
    (2)  edge node [left]  {restrict} (3)
    (3)  edge node [right] {interpolate} (4)
    (5)  edge node [left]  {restrict} (6)
    (6)  edge node [right] {interpolate} (7)
    (7)  edge node [right] {interpolate} (8)

    (9)  edge node [left]  {restrict} (10)
    (10) edge node [right] {interpolate} (11)
    (12) edge node [left]  {restrict} (13)
    (13) edge node [right] {interpolate} (14);
  \path[line,dashed]
    (4)  edge node [above] {K cycle} (5)
    (11) edge node [above] {$i$ times} (12);
  \path[line,dotted]
    (20) edge node [above] {} (1)
    (21) edge node [above] {} (8)
    (22) edge node [above] {} (9)
    (23) edge node [above] {} (11)
    (24) edge node [above] {} (14)
    (30) edge node [above] {} (3)
    (31) edge node [above] {} (6)
    (32) edge node [above] {} (10)
    (33) edge node [above] {} (13)
    (40) edge node [above] {} (4)
    (41) edge node [above] {} (7);
\end{tikzpicture}}
  \end{center}
  \caption{Illustration of Algorithm~\ref{alg:setup_iter} as iteration $i$ of the iterative setup phase. \label{fig:3lvlsetup}}
\end{figure}

\section{Numerical Results}\label{sec:results}
In this section we show extensive numerical results for our multilevel
domain decomposition adaptive algebraic multigrid
method (DD-$\alpha$AMG), especially for the three- and four-level
setting. As its predecessor in~\cite{Frommer:2013fsa} the method is
implemented as a parallel program in \texttt{C} using \texttt{MPI},
and we compute and apply the DD-$\alpha$AMG
preconditioner in single precision only. The outer
FGMRES iteration remains in double precision. 
The system on the coarsest level is solved via odd-even preconditioned GMRES to 
a given relative accuracy $\epsilon$.
The operators on all levels, including the finest-level Wilson Dirac
operator, are implemented as proposed in~\cite{Krieg:2010zz}. That is,
the underlying lattice is used to optimize the matrix vector multiplication.

As a new feature of the implementation, processes are now allowed to idle on the coarser
levels. This is necessary since a high degree of parallelization can
cause a lack of lattice sites on the coarser levels. In our
implementation we assume that sites which belong to a common aggregate
always share the same process. This allows to perform the
computational part of interpolation and restriction without
communication similarly to what has been done in the two-level approach. 

\begin{table}
\centering\scalebox{0.9}{\begin{tabular}{llcc}
\toprule
 & parameter                                 &         & default \\
\midrule 
setup & number of iterations                    & $\ninv$ & $6$ \\ 
 & number of test vectors                    & $N$     & $20$     \\
 & size of lattice-blocks for aggregates on level $1$  &         & $4^4$    \\
 & size of lattice-blocks for aggregates on level $\ell, \; \ell>1$  &     & $2^4$    \\
 & coarse system relative residual tolerance &         & \\
 & (stopping criterion for the coarse system)$^{(*)}$ &   $\epsilon$     &  $5\cdot10^{-2}$  \\
\midrule
solver & restart length of FGMRES            & $n_{kv}$& $10$ \\
 & relative residual tolerance  (stopping criterion)             & $\mathit{tol}$ & $10^{-10}$  \\ 
\midrule
smoother & number of pre-smoothing steps$^{(*)}$      & $\mu$   & $0$ \\
 & number of post-smoothing steps$^{(*)}$      & $\nu$   & $5$ \\
 & size of lattice-blocks in SAP$^{(*)}$                &         & $2^4$ \\
 & number of Minimal Residual (MR) iterations to  &    &  \\
 & solve the local systems in SAP$^{(*)}$& & $3$ \\
\midrule
K-cycle & maximal length$^{(*)}$              & & $5$ \\
	& maximal restarts$^{(*)}$            & & $2$ \\
	& relative residual tolerance (stopping criterion)$^{(*)}$ & & $10^{-1}$ \\
\bottomrule
  \end{tabular}}
  \caption{Parameters for the DD-$\alpha$AMG multi-level method.
   $(*):$ same in solver and setup.}
  \label{table:allparms}
\end{table}

Table~\ref{table:allparms} summarizes the default parameters used in
our experiments. Note that these parameters are the same as those used for the two-level
results in~\cite{Frommer:2013fsa} except that we reduced the restart length of the outer FGMRES
routine to $n_{kv}=10$ due to memory limitations when using a small number of cores.

\begin{table}
\centering\scalebox{0.9}{\begin{tabular}{ccccccc}
\toprule
id                                       &  lattice size            & pion mass     & CGNR      & shift         & clover          & provided by            \\
&  $N_t \times N_s^3$      & $m_\pi$ [MeV]& iterations & $m_0$         & term $c_\mathit{sw}$   &                 \\
\midrule
\ref{BMW_48_48}\conflabel{BMW_48_48}     &  $48 \times 48^3$        & $135$   & $53,\!932$ & $-0.09933$    & $1.00000$         & BMW-c~\cite{Durr:2010aw, BMW1}         \\
\ref{BMW_64_64}\conflabel{BMW_64_64}     &  $64 \times 64^3$        & $135$   & $84,\!207$ & $-0.05294$    & $1.00000$         & BMW-c~\cite{Durr:2010aw, BMW1}         \\
\ref{CLS_128_64}\conflabel{CLS_128_64}   &  $128 \times 64^3$       & $270$   & $45,\!804$ & $-0.34262$    & $1.75150$       & CLS~\cite{wwwCLS,Fritzsch:2012wq} \\
\ref{CLS_128_64_2}\conflabel{CLS_128_64_2} &  $128 \times 64^3$     & $190$   & $88,\!479$ & $-0.33485$    & $1.90952$       & CLS~\cite{wwwCLS,Fritzsch:2012wq} \\
\bottomrule
\end{tabular}}
\caption{Configurations used together with their parameters.
  For details about their generation we refer to the references.
  Pion masses rounded to steps of $5$ MeV.}
\label{table:allconfs}
\end{table}

In Table~\ref{table:allconfs} we give an overview of the
configurations used in our tests. The iteration count of CGNR, i.e., CG applied to the system
$D^\dagger D z = D^\dagger b$ with the residual $r = b-Dz$, can be used as an
indicator for the conditioning of the respective operator. All of the
configurations we use in our tests correspond to some of the most
challenging linear systems encountered in state-of-the-art lattice QCD calculations. 

In what follows we explore the potential benefits of additional levels
in the DD-$\alpha$AMG method. Special focus is put on the consequences of
additional levels in terms of the degree of parallelization, i.e., the size of the
local lattice kept on each node.
We tested the performance of the DD-$\alpha$AMG method with
different numbers of levels for a variety of cost measures
and analyzed the scaling behavior as a function of the bare mass $m_0$.
Finally, we compare the DD-$\alpha$AMG approach to
the recently improved version of the inexact deflation approach which now allows for 
inexact projection~\cite{wwwOPENQCD,Luescher2007}.

All results were obtained on the Juropa machine at J\"ulich
Supercomputing Centre, a cluster with $2,\!208$ compute nodes, each
with two Intel Xeon X5570 (Nehalem-EP) quad-core processors.
This machine provides a maximum of $8,\!192$ cores for a single job.
For compilation we used the \texttt{icc}-compiler with the optimization flags
\texttt{-O3}, \texttt{-ipo}, \texttt{-axSSE4.2} and \texttt{-m64}.

\subsection{Parallel Multilevel Methods} \label{sec:sweetspot}
Our first tests analyze the influence
of the degree of parallelization on the performance of the two- and
three-level method. 
This is an important aspect since using many processes
causes idle times on the coarser levels. In order to reduce or
even avoid these it might be promising to choose a
lower degree of parallelization. This helps to even out the
work-load on all levels such that the communication overhead
can be mostly neglected. In this manner we aim at an overall optimal 
use of resources, measured in core-minutes (time
to solution multiplied with the number of cores used), rather than just time to solution.

We tracked the performance of the two-
and three-level DD-$\alpha$AMG method for configurations~\ref{BMW_48_48} and~\ref{BMW_64_64} using
different degrees of parallelization, reported as the size of
the local lattice on each core. Our goal is to find the sweet spot with
respect to consumed core minutes.

\begin{figure}[htb]
\centering\scalebox{0.75}{\input{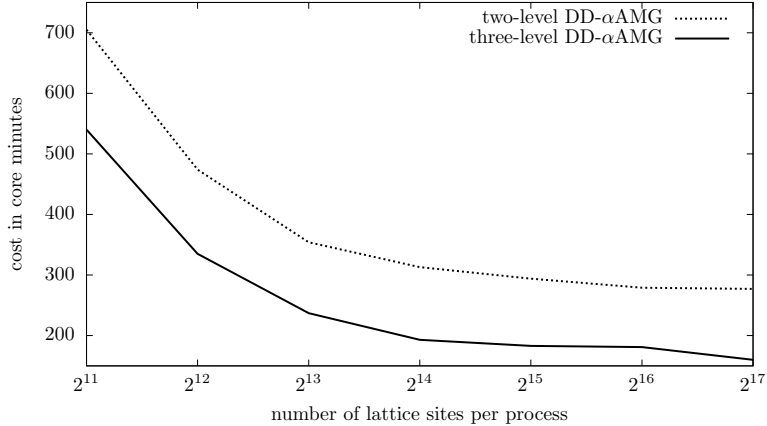}}
  \caption{Estimation of the sweet spot on configuration~\ref{BMW_48_48}.}
  \label{plot:sweet_spot48}
\end{figure}

Figure~\ref{plot:sweet_spot48} plots the dependence of the
solver performance on the
degree of parallelization for the rather small configuration~\ref{BMW_48_48}
with lattice size $48^4$.
Due to memory restrictions the minimal number of processors that can
be used is $36$ corresponding to roughly $2^{17}$ lattice sites per
processor. On the other end of the horizontal axis we are limited to
$5,\!184$ processors which corresponds to $2^{10}$ lattice sites per process on level~$1$.

Figure~\ref{plot:sweet_spot48} shows that as soon as a local lattice with
fewer than $8^4= 2^{12}$ lattice sites per processors on level $1$ is reached,
the performance of the two-level method exceeds the performance of the
three-level method. This is partly to be
expected, since this is exactly the spot where idling processors on the
second level cannot be avoided. To be more precise, on level
two we need at least $4 \times 2^3$ lattice sites per processor in
order to be able to apply SAP since we need one block of each color on each non-idling process.
On level three, i.e., the coarse level, we also assume to have at least two lattice sites
per non-idling process because we solve this system using odd-even preconditioning. Thus every second process
is idling on level two and three. Similarly, for $2^{11}$ local
sites three out of four processes 
and for $2^{10}$ sites $7$ out of $8$ processes are idling on levels two and three.
Thus, for the relatively small configuration~\ref{BMW_48_48} the
additional third level is advisable only if a relatively small degree of parallelism is used,
and even then the gain over the two-level method is rather modest.

\begin{figure}[htb]
\centering\scalebox{0.75}{\input{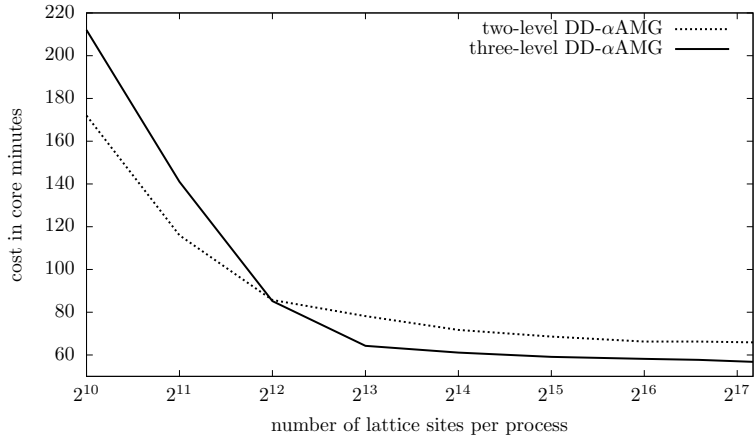}}
  \caption{Estimation of the sweet spot on configuration~\ref{BMW_64_64}.}
  \label{plot:sweet_spot}
\end{figure}

The picture changes, however, when we investigate the same dependence
for a larger configuration. In Figure~\ref{plot:sweet_spot} we see
an almost constant absolute gain when going from the two-level to the
three-level DD-$\alpha$AMG method. For any of the
considered degrees of parallelization, the three-level method outperforms the two
level method, and the
sweet spot is taken again for $2^{17}$ lattice sites per process
on the finest level, i.e., for $128$ processes and a $32  \times 16^3$ local
lattice on each process. For this particular case the three-level method
shows a speed up factor of $1.7$ over the two-level method. At the opposite end 
with a local lattice size of $2^{11}$ and $8,\!192$ processes, the three-level 
method still gains a factor of about $1.3$.

\subsection{Two, Three and Four Levels} \label{sec:2LVLvsMLVL}
Now we compare DD-$\alpha$AMG methods with two, three and four levels for all four
configurations, using only small numbers of cores, i.e., working
with large local lattices.
 
\begin{table}[htb]
\centering\scalebox{0.9}{\begin{tabular}{rlcccc}
\toprule
              & configuration     & \ref{BMW_48_48}  & \ref{BMW_64_64}  & \ref{CLS_128_64}  & \ref{CLS_128_64_2} \\
              & lattice size      & $48 \times 48^3$ & $64 \times 64^3$ & $128 \times 64^3$ & $128 \times 64^3$  \\
              & pion mass $m_\pi$ & $135\,$MeV       & $135\,$MeV       & $270\,$MeV        & $190\,$MeV         \\
\midrule
 two levels   & setup time        & $316$s           & $736$s           & $630$s            & $701$s             \\
              & solve time        & $48.6$s          & $130$s           & $113$s            & $141$s             \\
	      & solve iter        & $23$             & $22$             & $24$              & $28$ \\
\midrule
 three levels & setup time        & $374$s           & $744$s           & $719$s            & $948$s             \\
              & solve time        & $42.6$s          & $75.2$s          & $74.0$s           & $79.0$s            \\
	      & solve iter        & $24$             & $21$             & $22$              & $24$ \\
\midrule
 four levels  & setup time        & --               & $806$s           & $755$s            & $1,\!004$s         \\
              & solve time        & --               & $79.8$s          & $75.7$s           & $79.1$s            \\
	      & solve iter        & --               & $22$             & $22$              & $24$ \\
\midrule
              & processes         & $81$             & $128$            & $256$             & $256$              \\
local lattice & level 1           & $16 \times 16^3$ & $32 \times 16^3$ & $32 \times 16^3$  & $32 \times 16^3$   \\
              & level 2           & $4 \times 4^3$   & $8 \times 4^3$   & $8 \times 4^3$    & $8 \times 4^3$     \\
              & level 3           & $2 \times 2^3$   & $4 \times 2^3$   & $4 \times 2^3$    & $4 \times 2^3$     \\
              & level 4           & --               & $2 \times 1^3$   & $2 \times 1^3$    & $2 \times 1^3$     \\
\bottomrule
  \end{tabular}}
  \caption{Comparison of DD-$\alpha$AMG with two, three and four levels for a small number of processes,
  parameters from Tables~\ref{table:allparms} and~\ref{table:allconfs}.}
\label{table:2vs3vs4}
\end{table}

Table~\ref{table:2vs3vs4}  shows that the three-level method outperforms the
other variants in terms of consumed core minutes in all tests except for configuration~\ref{BMW_48_48}
by factors between $1.5$ and $1.7$, and it also outperforms the four-level method.
Note that the time spent in the setup naturally increases when using
additional levels, since additional operators need to be set up 
and only in one test this additional work amortized immediately in one
solve. While we still believe that we can improve the setup routine
and with it the additional overhead to be paid for additional levels,
the overhead has to be kept in mind when choosing the number of levels,
depending on the number of right-hand-sides to be solved. Interpreting 
Table~\ref{table:2vs3vs4} with respect to the system size, we see that the performance gain of
additional levels grows with the system size. This indicates that we can expect 
the gain of three- or potentially four-level approaches to grow for even larger lattices.

\begin{table}[htb]
\centering\scalebox{0.9}{\begin{tabular}{rlccccc}
\toprule
              & configuration     & \multicolumn{2}{c}{\ref{BMW_48_48}}  &\quad\quad&  \multicolumn{2}{c}{\ref{BMW_64_64}} \\
              & lattice size      & \multicolumn{2}{c}{$48 \times 48^3$} &&  \multicolumn{2}{c}{$64 \times 64^3$} \\
              & pion mass $m_\pi$ &  \multicolumn{2}{c}{$135\,$MeV} &&  \multicolumn{2}{c}{$135\,$MeV} \\
\midrule
              & levels            & $2$              & $3$              && $2$              & $3$              \\
              & setup time        & $14.9$s          & $40.0$s          && $24.5$s          & $46.1$s          \\
	      & solve time        & $2.69$s          & $3.26$s          && $5.21$s          & $3.95$s          \\
              & solve iter        & $23$             & $24$             && $23$             & $22$             \\
\midrule
level $1$     & consumed time     & $0.880$s         & $0.895$s         && $1.04$s          & $0.930$s         \\
              & wait time         & $0.110$s         & $0.115$s         && $0.124$s         & $0.135$s         \\
level $2$     & consumed time     & $1.81$s          & $1.50$s          && $4.17$s          & $1.48$s          \\
              & wait time         & $0.0972$s        & $0.0557$s        && $0.171$s         & $0.0430$s        \\
level $3$     & consumed time     & --               & $0.865$s         && --               & $1.54$s          \\
              & wait time         & --               & $0.0993$s        && --               & $0.0942$s        \\
summarized    & total wait time   & $0.207$s         & $0.270$s         && $0.295$s         & $0.272$s         \\
\midrule
	      & processes         & $2,\!592$        & $2,\!592$        && $8,\!192$        & $8,\!192$        \\
local lattice & level 1           & $4 \times 8^3$   & $4 \times 8^3$   && $4 \times 8^3$   & $4 \times 8^3$   \\
              & level 2           & $1 \times 2^3$   & $4 \times 2^3$   && $1 \times 2^3$   & $4 \times 2^3$   \\
	      & level 3           & --               & $2 \times 1^3$   && --               & $2 \times 1^3$   \\
\bottomrule
  \end{tabular}}
  \caption{Comparison of DD-$\alpha$AMG with two and three levels for a large number of processes,
  parameters from Tables~\ref{table:allparms}  and~\ref{table:allconfs}.}
\label{table:2vs3}
\end{table}

In Table~\ref{table:2vs3} we show additional timings for
comparatively large numbers of processes where idle times on the coarser
levels occur. More precisely, for the number of processes chosen in
the tests, three out of four processes idle on the second and third
level within the three-level method while for the two-level variant
there are no idle times at all.
For configuration~\ref{BMW_48_48} we observe that the two-level method outperforms
the three-level method with respect to setup and solve time as
expected based on Figure~\ref{BMW_48_48}. In this particular case the three-level
method is unable to transfer enough work to the coarse level and level
two remains the most expensive part of the solve phase.
Together with the fact that for this configuration the second level can be solved quite
efficiently by odd-even preconditioned GMRES alone, a third level does not pay off.

For the larger configuration~\ref{BMW_64_64}, the situation is similar
with respect to setup timings, but the second level
seems to be much more ill-conditioned which yields an advantage for the three-level method
regarding the solve time.
Thus in situations where the increased setup time can be compensated for by solving
systems with several right-hand-sides, the three-level method pays off.

Besides these observations we also note that the influence of wait times caused by nearest neighbor
communication on the coarse level even with a local lattice size of $2 \times 1^3$
can be neglected. Thus the overall performance loss on the coarser levels
is dominated by the percentage of idling processes and not
the communication overhead.

\subsection{Comparison with CGNR}
We now want to study the two- and three-level method in more detail,
namely, with respect to absolute cost measures and in comparison to
the conventional Krylov subspace method of choice, CGNR. We choose to
show absolute cost measures, e.g., flop count and core minutes rather than
wall-clock time. The performance of both methods, DD-$\alpha$AMG and CGNR, in terms of flop/s
(floating point operations per second) can differ dramatically
depending on the level of optimization of the Dirac
operator and on the degree of parallelization chosen, whereas core minutes 
represent a measure that takes both important resources, time and hardware, 
into account.

In Table~\ref{table:3LVLvsCGNR} we show results of our method and CGNR
for configuration~\ref{BMW_64_64}. First note that compared to CGNR
the two-level method speeds up the calculation by an order of
magnitude and cuts down the flop count per lattice site
and overall core minute count even by more than two orders of magnitude. The
addition of a third level yields another factor of roughly $1.7$
across the board with respect to the two-level method.
It is noteworthy that the three-level DD-$\alpha$AMG
performs at $2.86$ Gflop/s per core, corresponding to $12.2\%$ peak performance, on Juropa with a
pure \texttt{C}-code, i.e., without any machine specific optimization.

The abysmal Gflop/s per core performance of CGNR is mainly due to two facts.
First we run CGNR completely in double precision, second we are limited to at most
$8,\!192$ cores per run and the corresponding local lattice size of $4
\times 8^3$ leads to a problem size exceeding the cache size.

\begin{table}[htb]
\centering\scalebox{0.9}{\begin{tabular}{rccc}
\toprule
                        & three-level      & two-level        &                     \\
                        & DD-$\alpha$AMG   & DD-$\alpha$AMG   & CGNR                \\
\midrule
processes               & $128$            & $128$            & $8,\!192$           \\
solve time              & $75.2$s          & $130$s           & $816$s              \\
consumed core minutes   & $160$            & $277$            & $111,\!446$         \\
consumed Mflop per site & $1.64$           & $2.97$           & $364$               \\
Gflop/s per core        & $2.86$           & $2.99$           & $0.91$              \\
\bottomrule
  \end{tabular}}
  \caption{Comparison of three-level DD-$\alpha$AMG with CGNR,
           parameters from Tables~\ref{table:allparms} and~\ref{table:allconfs}.}
\label{table:3LVLvsCGNR}
\end{table}

Combined with the results reported in Section~\ref{sec:2LVLvsMLVL} our results
suggest that in situations where total run-time is not of great
importance and where several Dirac systems with different right-hand-sides must be solved, e.g., in configuration analysis, it is best to run a
multilevel method with a low degree of parallelization. This gives the largest number of
system solves per core minute, i.e., the largest data per Euro
ratio. On the other hand, in a situation,
where total run-time is of utmost importance, e.g., the generation of
configurations within the HMC process, it might be advisable to use only a two-level
method and a high degree of parallelization. 

\subsection{Scaling Tests}\label{sec:scaling_tests}
An important motivation for the development of multilevel preconditioners for the Wilson
Dirac system has always been the removal of ``critical slowing down'',
i.e., the observed dramatic drop in performance when the mass parameter
approaches its critical value. In the next set of tests we report
the time to solution of the two-, three- and four-level DD-$\alpha$AMG
method with respect to the bare
mass $m_0$ for configuration~\ref{BMW_64_64}. These tests are again
carried out using only $128$ cores, i.e., a low degree of parallelization.

\begin{figure}[htb]
\centering\scalebox{0.75}{\input{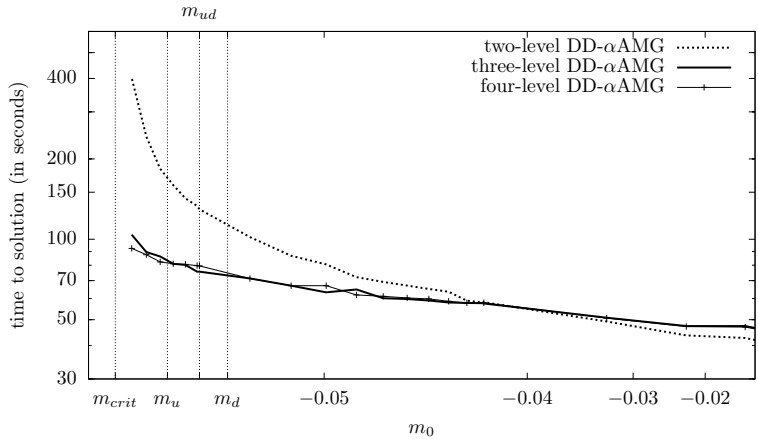}}
  \caption{Scaling of DD-$\alpha$AMG with the bare mass
    $m_0$. In here, $m_{ud} = -0.05294$ denotes the physical mass
    parameter for which the configuration was thermalized and
$m_{crit} = -0.05419$~\cite{Durr:2010aw, BMW1}
is the critical mass.}
  \label{plot:mass_scaling}
\end{figure}

We see in Figure~\ref{plot:mass_scaling} that the time
to solution of the two-level method increases much 
more rapidly than the time to solution of the three-level method when approaching the 
critical bare mass $m_{crit}$.
Further we see that beyond $m_u$ the four-level method starts to 
outperform the three-level method.
Note that for the computation of observables in lattice QCD
it might actually be necessary to solve the Dirac equation for mass parameters
beyond $m_{ud}$, e.g, to account for the up-down quark mass 
difference~\cite{Borsanyi:2013lga,Finkenrath:2013soa}.
For configuration~\ref{BMW_64_64} this requires a mass parameter
$m_u = -0.05347$ which leads to a factor of approximately $1.6$
in the condition number compared to $m_{ud}$.

\begin{figure}[htb]
\centering\scalebox{0.75}{\input{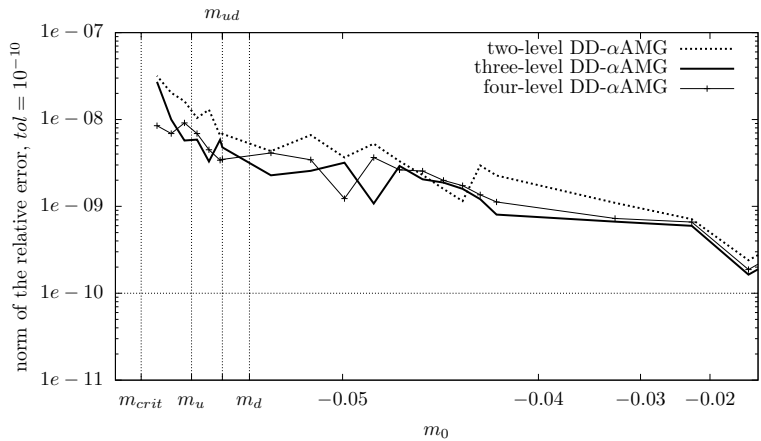}}
  \caption{Scaling of the remaining error obtained from DD-$\alpha$AMG}
  \label{plot:error_scaling}
\end{figure}

We also tracked the norm of the relative error $||e||/||z^{*}||=||z^{*}-z||/||z^{*}||$
as a function of $m_0$ for a pre-determined solution $z^{*}$.
Figure~\ref{plot:error_scaling} shows that the error 
only slightly increases in the range
of the physical masses when using more levels even though the tolerance
for the K-cycle on the second level is a factor of $2$ less precise
than for odd-even GMRES within the two-level method. When approaching the
critical bare mass, four-level DD-$\alpha$AMG also provides the
most stable error.

\subsection{Comparison with Inexact Deflation with Inexact Projection}
Recently the implementation of inexact deflation was upgraded within
the openQCD code~\cite{wwwOPENQCD}. The new version of inexact
deflation, termed ``inexact deflation with inexact projection'' is similar 
in spirit to our method proposed in~\cite{Frommer:2013fsa}, while it 
differs in its construction of the interpolation and
the coarse-level operator. In the inexact deflation
approach $\gamma_5$-symmetry is not preserved on the coarse
level. 

In order to account for this recent upgrade of the openQCD code, we
compare it with multilevel DD-$\alpha$AMG. As
the openQCD code uses open boundary conditions in time direction we
cannot use our set of configurations with this code. Thus we
integrated the new modules containing the inexact deflation with inexact projection method
into the DD-HMC~\cite{wwwDDHMC} and compared both methods.

\begin{table}[htb]
\centering\scalebox{0.9}{\begin{tabular}{rcccc}
\toprule
                     & three-level    & two-level      & inexact deflation & inexact deflation \\
  & DD-$\alpha$AMG & DD-$\alpha$AMG & (DD-HMC-1.2.2)~\cite{wwwDDHMC} & with inexact projection~\cite{wwwOPENQCD} \\
\midrule
test vectors $N$     & $20$           & $20$           & $20$              & $32$              \\
setup bootstrap iter & $6$            & $6$            & $10$              & $6$               \\
SAP block size       & $2^4$          & $2^4$          & $4^4$             & $4^4$             \\
block solver iter    & $3$            & $3$            & $4$               & $4$               \\
\midrule
setup time           & $744$s         & $736$s         & $681$s            & $897$s            \\
solve iter           & $21$           & $22$           & $48$              & $18$              \\
solve time           & $75.2$s        & $130$s         & $223$s            & $96.6$s           \\
\bottomrule
  \end{tabular}}
  \caption{Comparison on configuration~\ref{BMW_64_64} using $128$ processes,
           parameters from Tables~\ref{table:allparms} and~\ref{table:allconfs}.}
\label{table:ID_comparison}
\end{table}

Table~\ref{table:ID_comparison} shows that the inexact deflation with
inexact projection method for configuration~\ref{BMW_64_64} is $2.3$
times faster than without inexact projection, and also a factor of
$1.35$ faster than two-level DD-$\alpha$AMG. This is mainly due
to the fact that inexact deflation with inexact projection does not preserve the $\gamma_5$-symmetry 
on the coarse level. It therefore ends up with only half as many variables and 
a four times less complex operator on the
coarse level if the same number of test vectors is used. As a
consequence we found numerically that the inexact 
deflation with inexact projection approach has to compute more test vectors, but not 
twice as many, to
construct a comparably ``rich'' coarse-level subspace which is needed
to achieve the same number of iterations as the $\gamma_{5}$-symmetry preserving approach.
Although the setup time of the three-level DD-$\alpha$AMG method is
slightly longer than that of two of the other methods, the reduced
solve time more than compensates for this already for a single
right-hand-side.

Note that the results presented so far for the $\gamma_5$-symmetry
preserving interpolation show that the recursive extension of
DD-$\alpha$AMG works and, in particular, that the SAP smoother still
works on coarse levels. It is unclear whether this is still the case if $\gamma_5$-symmetry
is not preserved on the coarse level, so that a recursive extension of this
approach might not benefit as much from additional levels.
These questions can probably only be answered experimentally.

\begin{figure}[htb]
\centering\scalebox{0.75}{\input{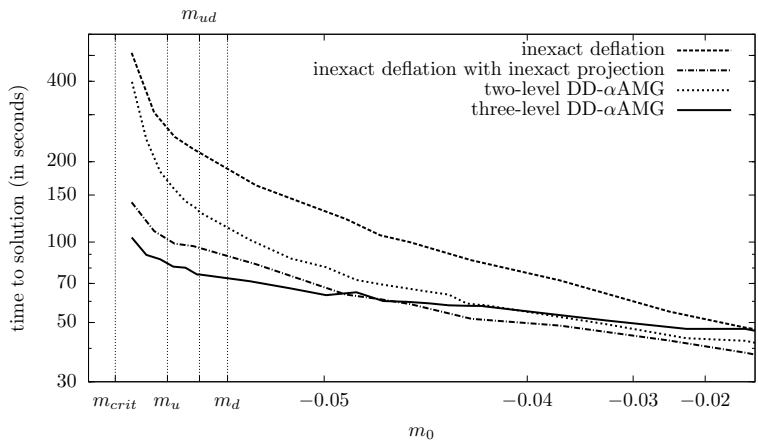}}
  \caption{Scaling with the bare mass $m_0$ of inexact deflation and DD-$\alpha$AMG on configuration~\ref{BMW_64_64} using $128$ processes.}
  \label{plot:ID_mass_scaling}
\end{figure}

As the behavior of solvers approaching the critical mass is an
important bench-mark, we also investigate the scaling of the new inexact deflation with
inexact projection method.
Figure~\ref{plot:ID_mass_scaling} shows the scaling behavior
of all the methods reported in Table~\ref{table:allparms}
as a function of the bare mass $m_0$. Inexact deflation with inexact projection and two-level DD-$\alpha$AMG
show a similar scaling behavior until $m_{ud}$. Beyond
this mass, two-level DD-$\alpha$AMG tends to scale similarly to the
ordinary inexact deflation approach. Besides that we note that the
upgrade of the inexact deflation method also leads to an improved
scaling behavior.
Thus the ability to solve the coarse system equations inexactly 
and the ability to use more test vectors and still having a cheap
coarse-level operator can fundamentally influence the scaling behavior
of the two-level method. Though the overall best scaling curve belongs
to three-level DD-$\alpha$AMG and the third level already pays off for
heavier masses than $m_d$ and will pay off even more in the future
when even larger lattices will be used.

\section{Conclusions and Outlook}
We successfully extended our two-level method to a true multilevel
method including a multilevel setup. For certain cases our
three-level implementation obtains speed-ups of up to a factor of $1.7$
compared to the two-level version for physical masses. The scaling
behavior with respect to the bare mass and the lattice size shows great potential
for future lattice QCD computations on even larger lattices.
Another factor of $1.5$ - $2$ could probably be obtained by machine
specific optimization (SSE/AVX/QPX). We are currently incorporating our
algorithm into the production codes of our collaborators within SFB/TRR55.
Furthermore we are planning to investigate how to incorporate the
DD-$\alpha$AMG method into the Hybrid Monte Carlo Method, i.e., updating
the multilevel hierarchy in a cost efficient way along the MD trajectory.

\section*{Acknowledgments}
We thank the Budapest-Marseille-Wuppertal collaboration and the CLS
consortium for providing configurations. All results were computed
on Juropa at J\"ulich Supercomputing Centre (JSC). We would also like
to acknowledge James Brannick (Pennsylvania State University) for his
advice regarding the development of our multilevel method, especially
the extension of the bootstrap setup to the QCD context. Further
thanks go to Kalman Szab{\'o} (Bergische Universit\"at Wuppertal)
for his support and counsel regarding the implementation of the
multilevel solver and Wolfgang S{\"o}ldner (University of Regensburg)
for his time spent discussing our code and helping out with the
I/O-interfaces.

\bibliographystyle{siam}
\bibliography{3lvl_note}

\end{document}